\documentclass{PoS}
\usepackage{array}

\title{Universality of TMD distribution functions}

\ShortTitle{Universality of TMD distribution functions}

\author{\speaker{Maarten BUFFING}\\
        Nikhef and Department of Physics and Astronomy, VU University Amsterdam\\
        De Boelelaan 1081, NL-1081 HV Amsterdam, the Netherlands\\
        E-mail: \email{m.g.a.buffing@vu.nl}}

\author{Asmita MUKHERJEE\\
        Department of Physics, Indian Institute of Technology Bombay\\
        Powai, Mumbai 400076, India\\
        E-mail: \email{asmita@phy.iitb.ac.in}}

\author{Piet MULDERS\\
        Nikhef and Department of Physics and Astronomy, VU University Amsterdam\\
        De Boelelaan 1081, NL-1081 HV Amsterdam, the Netherlands\\
        E-mail: \email{mulders@few.vu.nl}}

\abstract{We introduce transverse momentum dependent parton distribution functions (TMDs) for gluons with definite rank. The rank refers to the azimuthal dependence corresponding to the tensorial structure in transverse momenta multiplying universal functions only depending on $x$ and $p_{\scriptscriptstyle T}^2$. In this way only a finite number of functions of definite rank remains for a target with the maximal rank depending on its spin. Gauge links, required for color gauge invariance, enter in the explicit description of the matrix elements corresponding to these TMDs and account for their process dependence. In this way a general gauge link dependent function is expressed in the universal set, where all process (i.e. gauge link) dependence is isolated in gluonic pole factors multiplying the universal TMDs of definite rank.}

\FullConference{XXI International Workshop on Deep-Inelastic Scattering and Related Subjects\\
                 22-26 April, 2013\\
                 Marseilles, France}

\begin{document}

\section{Introduction}
\begin{figure}[!b]
\begin{center}
\includegraphics[width=.38\textwidth]{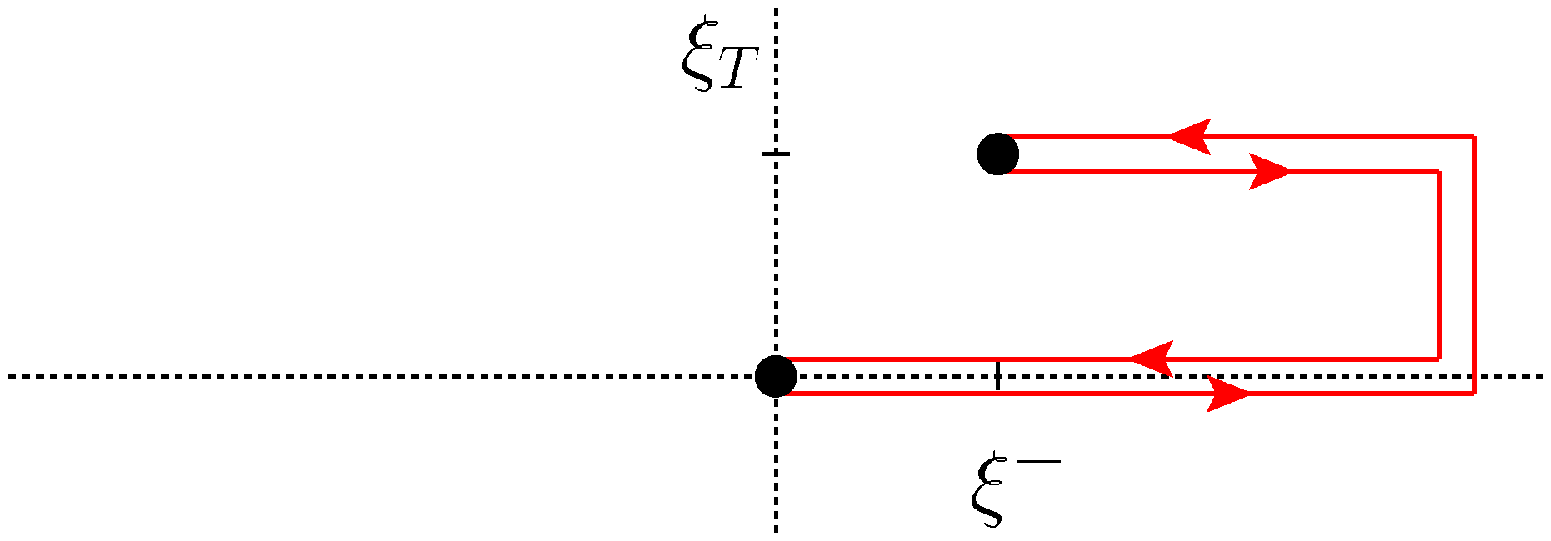}
\hspace{12mm}
\includegraphics[width=.38\textwidth]{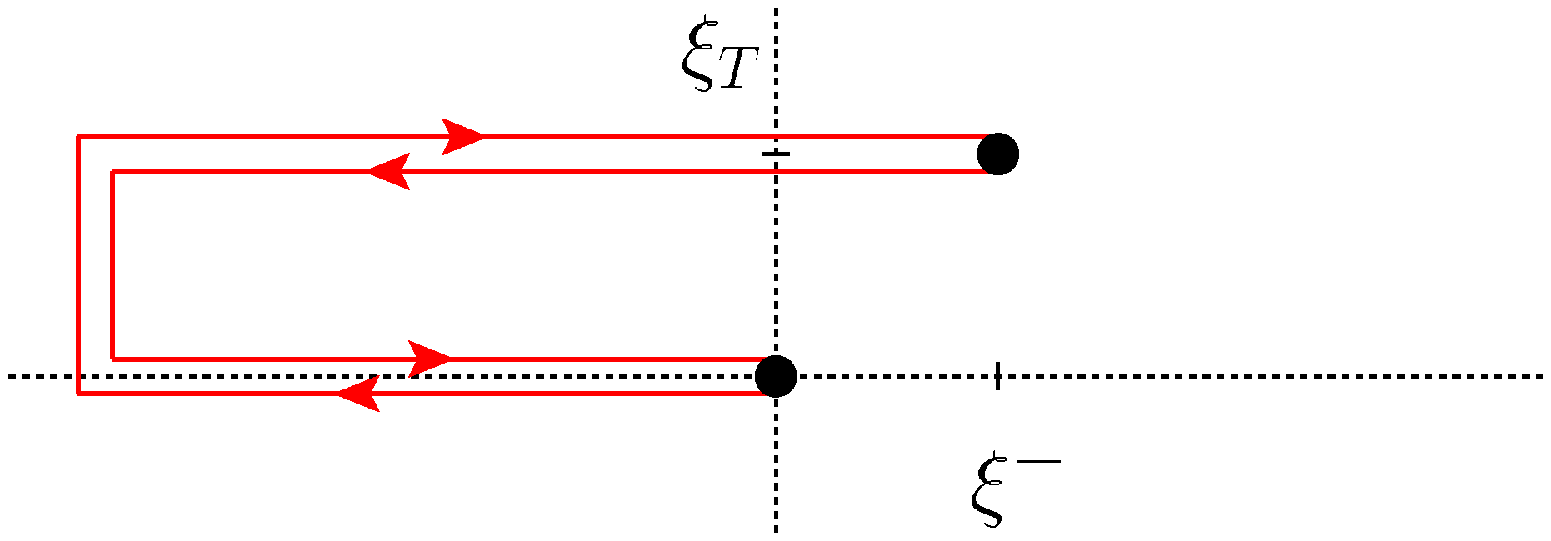}
\\
(a)\hspace{72mm} (b)
\\[3mm]
\includegraphics[width=.38\textwidth]{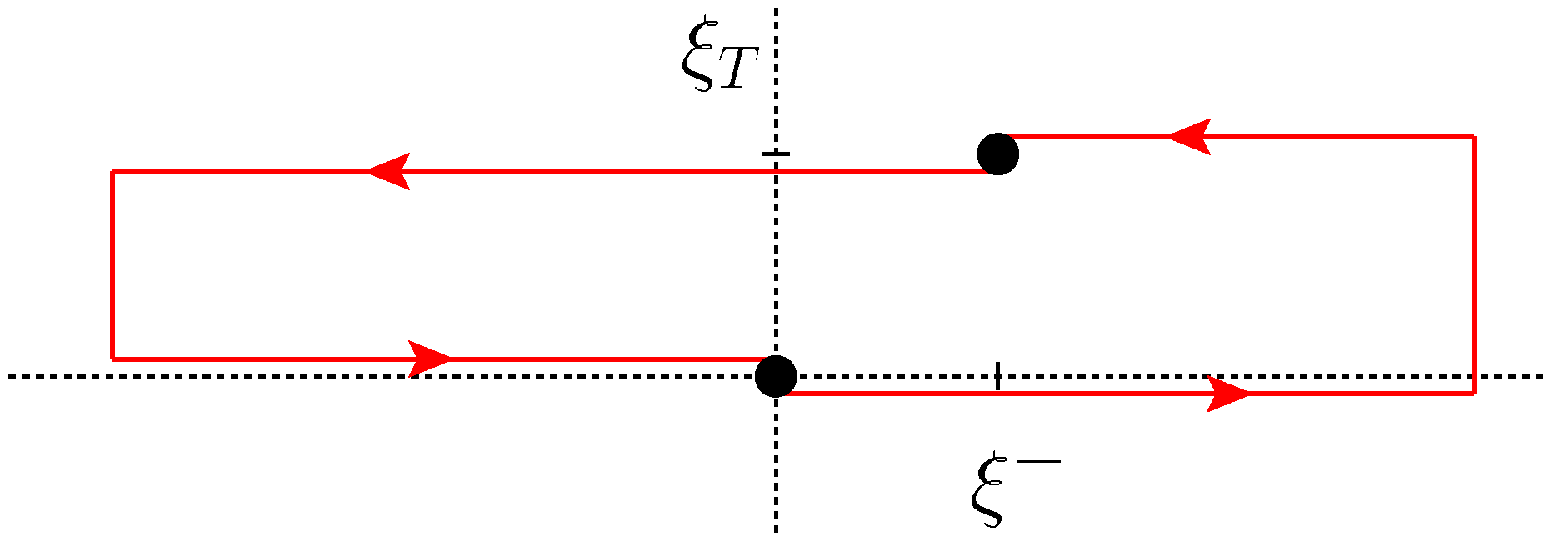}
\hspace{12mm}
\includegraphics[width=.38\textwidth]{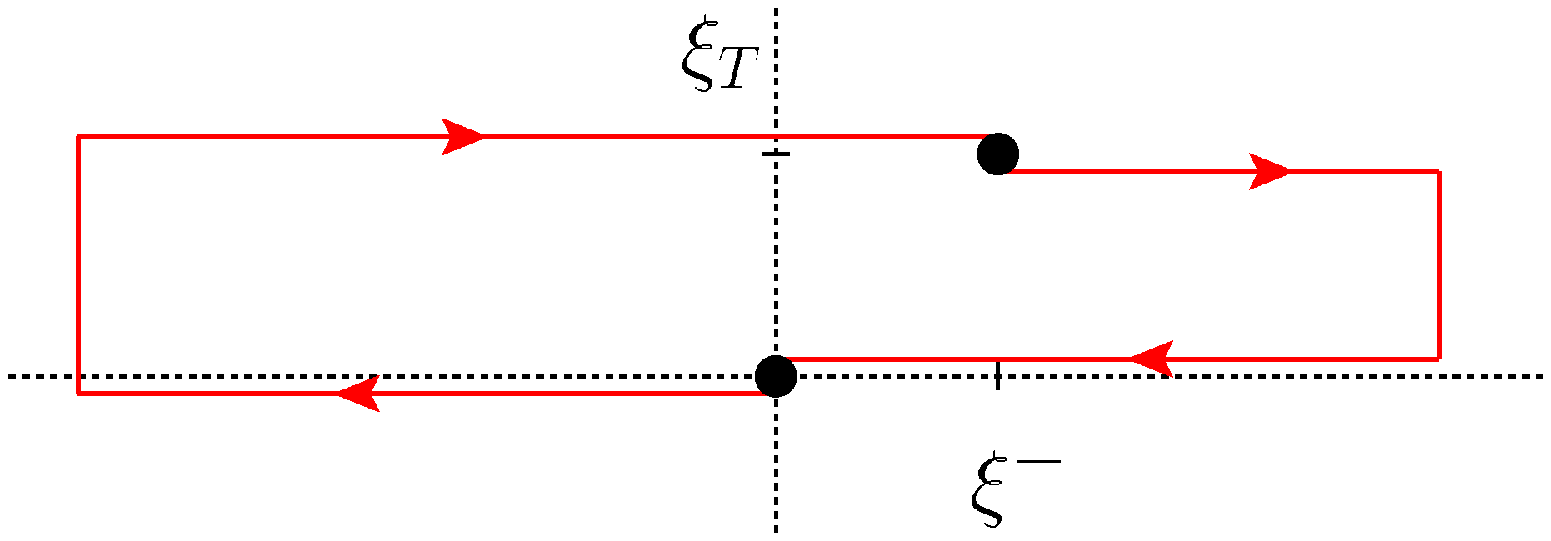}
\\
(c)\hspace{72mm} (d)
\caption{The gauge link structures (a) $[+,+^\dagger]$, (b) $[-,-^\dagger]$, (c) $[+,-^\dagger]$ and (d) $[-,+^\dagger]$.}
\label{f:GL++--}
\end{center}
\end{figure}

We start from the gluon distribution function, which can be written in the form of a matrix element as~\cite{Mulders:2000sh,Bomhof:2006dp,Bomhof:2007xt}
\begin{equation}
\Gamma^{[U,U^\prime]\,\mu\nu}(x,p_{\scriptscriptstyle T};n) ={\int}\frac{d\,\xi{\cdot}P\,d^2\xi_{\scriptscriptstyle T}}{(2\pi)^3}\ e^{ip\cdot\xi}\,\langle P{,}S|\,F^{n\mu}(0)\,U_{[0,\xi]}^{\phantom{\prime}}\,F^{n\nu}(\xi)\,U_{[\xi,0]}^\prime\,|P{,}S\rangle\biggr|_{\textrm{\scriptsize{LF}}}.
\label{e:matrix}
\end{equation}
The gauge links $U_{[0,\xi]}^{\phantom{\prime}}$ and $U_{[\xi,0]}^\prime$, which are path ordered exponentials, are needed to make the correlator gauge invariant. Depending on the process under consideration different gauge links will appear.
A number of gauge link structures exist, denoted as $[U,U^\prime]$, connecting the positions $0$ and $\xi$ in different ways. As basic building blocks we use the staple links $U_{[0,\xi]}^{[\pm]} = U_{[0,\pm\infty]}^{[n]} U_{[0_{\scriptscriptstyle T},\xi_{\scriptscriptstyle T}]}^{{\scriptscriptstyle T}}U_{[\pm \infty,\xi]}^{[n]}$. The simplest combinations allowed for $[U,U^\prime]$ are $[+,+^\dagger]$, $[-,-^\dagger]$, $[+,-^\dagger]$ and $[-,+^\dagger]$, which are illustrated in Fig.~\ref{f:GL++--}. More complicated possibilities, e.g. with additional (traced) Wilson loops of the form $U^{[\square]}=U_{[0,\xi]}^{[+]}U_{[\xi,0]}^{[-]}$ = $U_{[0,\xi]}^{[+]}U_{[0,\xi]}^{[-]\dagger}$ or $U^{[\square]^{\dagger}}$ = $U_{[0,\xi]}^{[-]}U_{[\xi,0]}^{[+]}$ = $U_{[0,\xi]}^{[-]}U_{[0,\xi]}^{[+]\dagger}$ are allowed as well. A list with all type of contributions can be found in Ref.~\cite{Buffing:2013kca}.
Since the above correlator cannot be calculated from first principles, an expansion in terms of TMD PDFs is used, which at the level of leading twist contributions is given by~\cite{Mulders:2000sh,Meissner:2007rx}
\begin{eqnarray}
2x\,\Gamma^{\mu\nu [U]}(x{,}p_{\scriptscriptstyle T}) &=& 
-g_T^{\mu\nu}\,f_1^{g [U]}(x{,}p_{\scriptscriptstyle T}^2)
+g_T^{\mu\nu}\frac{\epsilon_T^{p_TS_T}}{M}\,f_{1T}^{\perp g[U]}(x{,}p_{\scriptscriptstyle T}^2)
\nonumber\\&&
+i\epsilon_T^{\mu\nu}\;g_{1s}^{g [U]}(x{,}p_{\scriptscriptstyle T})
+\bigg(\frac{p_T^\mu p_T^\nu}{M^2}\,{-}\,g_T^{\mu\nu}\frac{p_{\scriptscriptstyle T}^2}{2M^2}\bigg)\;h_1^{\perp g [U]}(x{,}p_{\scriptscriptstyle T}^2)
\nonumber\\ &&
-\frac{\epsilon_T^{p_T\{\mu}p_T^{\nu\}}}{2M^2}\;h_{1s}^{\perp g [U]}(x{,}p_{\scriptscriptstyle T})
-\frac{\epsilon_T^{p_T\{\mu}S_T^{\nu\}}{+}\epsilon_T^{S_T\{\mu}p_T^{\nu\}}}{4M}\;h_{1T}^{g[U]}(x{,}p_{\scriptscriptstyle T}^2).
\label{e:GluonCorr}
\end{eqnarray}
We have used that $S^\mu = S_{\scriptscriptstyle L}P^\mu + S^\mu_{\scriptscriptstyle T} + M^2\,S_{\scriptscriptstyle L}n^\mu$. For $g_{1s}^{g [U]}$ and $h_{1s}^{\perp g [U]}$ the shorthand notation
\begin{equation}
g_{1s}^{g [U]}(x,p_{\scriptscriptstyle T})=S_{\scriptscriptstyle L} g_{1L}^{g [U]}(x,p_{\scriptscriptstyle T}^2)-\frac{p_{\scriptscriptstyle T}\cdot S_{\scriptscriptstyle T}}{M}g_{1T}^{g [U]}(x,p_{\scriptscriptstyle T}^2)
\end{equation}
is used.
The gauge link dependence on the rhs of Eq.~\ref{e:GluonCorr} is hidden in the TMDs, as a result of which the TMDs themselves are potentially process dependent. Note that $f_{1T}^{\perp g}$, $h_{1T}^{g}$, $h_{1L}^{\perp g}$ and $h_{1T}^{\perp g}$ are naive T-odd.

\section{Formalism}
The universality breaking of the TMDs is troublesome at least, due to its implications that it is no longer possible to use TMDs measured in one process for describing another process. In order to save some predictability, first of all it has to be decided how to define the TMDs and what mechanism is used for identifying them from a theoretical point of view. For this, we use transverse moments, defined as
\begin{equation}
\Gamma_{\partial\ldots\partial}^{\alpha_1\ldots\alpha_n[U]}(x)\equiv \int d^2p_{\scriptscriptstyle T}\ p_{\scriptscriptstyle T}^{\alpha_1}\ldots p_{\scriptscriptstyle T}^{\alpha_n}\,\Gamma^{[U]}(x,p_{\scriptscriptstyle T}), 
\end{equation}
which are weightings with transverse momenta $p_{\scriptscriptstyle T}$. Since these transverse momenta become derivatives in coordinate space, they are sensitive to all objects that depend on the transverse coordinate $\xi_T$, including gauge links~\cite{Buffing:2011mj,Buffing:2012sz}. As an example, we consider the weighting with one factor of $p_{\scriptscriptstyle T}$ and refer to Ref.~\cite{Buffing:2013kca} for the details regarding higher transverse weightings.

For single transverse weighting of the matrix element in Eq.~\ref{e:matrix} we end up with the result~\cite{Bomhof:2007xt,Buffing:2013kca}
\begin{eqnarray}
\Gamma_{\partial}^{\alpha [U]}(x)\equiv\int d^2p_{\scriptscriptstyle T}\ p_{\scriptscriptstyle T}^{\alpha}\,\Gamma^{[U]}(x,p_{\scriptscriptstyle T})=\widetilde\Gamma_{\partial}^{\alpha}(x)+C_{G,1}^{[U]}\,\Gamma_{G,1}^{\alpha}(x)+C_{G,2}^{[U]}\,\Gamma_{G,2}^{\alpha}(x).
\label{e:moment1}
\end{eqnarray}
In this, the indices $\partial$ and $G$ in the correlators indicate the operator structure of the partonic operators that appear due to the weighting. The correlators in Eq.~\ref{e:moment1} have as basis
\begin{eqnarray}
&&\Gamma_{D}^{\mu\nu,\alpha[U]}(x,x-x_1)=\int \frac{d\,\xi{\cdot}P}{2\pi}\frac{d\,\eta{\cdot}P}{2\pi}\ e^{ix_1 (\eta\cdot P)}e^{i (x-x_1)(\xi \cdot P)} \nonumber \\
&&\hspace{30mm} \times\textrm{Tr}\langle P,S|F^{n\mu}(0)\big[U_{[0,\eta]}^{[n]}
iD_{\scriptscriptstyle T}^{\alpha}(\eta)U_{[\eta,0]}^{[n]},U_{[0,\xi]}^{[n]}F^{n\nu}(\xi)U_{[\xi,0]}^{[n]}\big]|P,S\rangle\Big|_{\textrm{\scriptsize{LC}}}, 
\label{e:GammaD} \\
&&\Gamma_{F,1}^{\mu\nu,\alpha[U]}(x,x-x_1)=\int \frac{d\,\xi{\cdot}P}{2\pi}\frac{d\,\eta{\cdot}P}{2\pi}\ e^{ix_1 (\eta\cdot P)}e^{i (x-x_1)(\xi \cdot P)} \nonumber \\
&&\hspace{30mm} \times\textrm{Tr}\langle P,S|F^{n\mu}(0)\big[U_{[0,\eta]}^{[n]}
F^{n\alpha}(\eta)U_{[\eta,0]}^{[n]},U_{[0,\xi]}^{[n]}F^{n\nu}(\xi)U_{[\xi,0]}^{[n]}\big]|P,S\rangle\Big|_{\textrm{\scriptsize{LC}}},
\label{e:GammaF1} 
\end{eqnarray}
see Ref.~\cite{Buffing:2013kca} for more details. Another correlator, $\Gamma_{F,2}^{\mu\nu,\alpha[U]}(x,x-x_1)$, is given by a similar expression as $\Gamma_{F,1}^{\mu\nu,\alpha[U]}(x,x-x_1)$ in Eq.~\ref{e:GammaF1}, with the commutator of the partonic gluon fields replaced by an anticommutator. The correlators as they appear in Eq.~\ref{e:moment1} are then given by
\begin{eqnarray}
&&\Gamma_{D}^{\mu\nu,\alpha}(x)=\int dx_1\ \Gamma_{D}^{\mu\nu,\alpha}(x,x-x_1), \label{e:GammaD1} \\
&&\Gamma_A^{\mu\nu,\alpha}(x)=\int dx_1 \ \textrm{PV}\frac{i}{x_1}\Gamma_{F,1}^{\mu\nu,\alpha}(x,x-x_1), \label{e:GammaA1} \\
&&\widetilde\Gamma_\partial^{\mu\nu,\alpha}(x) \equiv \Gamma_{D}^{\mu\nu,\alpha}(x)-\Gamma_{A}^{\mu\nu,\alpha}(x), \label{e:Gammad1} \\
&&\Gamma_{G,c}^{\mu\nu,\alpha}(x)=\Gamma_{F,c}^{\mu\nu,\alpha}(x,x) \label{e:GammaGc1}.
\end{eqnarray}
In our notation, the correlator with index $\partial$ corresponds to a operator structure containing a commutator of the partial derivative operator combination $i\partial_{\scriptscriptstyle T}=iD_{\scriptscriptstyle T}-A_{\scriptscriptstyle T}$ with the gluon field $F(\xi)$. The correlators with the index $G$ are gluonic pole contributions. These gluonic pole correlators are multiparton correlators with zero momentum gluons~\cite{Efremov:1981sh,Efremov:1984ip,Qiu:1991pp,Qiu:1991wg,Qiu:1998ia,Kanazawa:2000hz} and are multiplied with calculable gluonic pole factors, which contain all the process dependence.

Furthermore, the index $c$ indicates the color structure of the operators in the correlator. Since the gluonic poles are combinations of gluon fields, depending on the gauge link structure they can occur with plus or minus signs, giving commutators and anticommutators of gluon fields. This implies that the gluonic pole can appear as a commutator ($c=1$) or anticommutator ($c=2$) with the field $F(\xi)$. Although this was already shown in e.g. Ref.~\cite{Bomhof:2007xt} for the single weighted case, we have generalized it to higher weightings, where more color structures are possible.

Transverse moments can be used for the correlator in terms of TMDs as well, where we define
\begin{equation}
f_{\ldots}^{g (m)}(x,p_{\scriptscriptstyle T}^2)=\left(\frac{-p_{\scriptscriptstyle T}^2}{2M^2}\right)^m\,f_{\ldots}^{g}(x,p_{\scriptscriptstyle T}^2). \label{e:tm}
\end{equation}
Performing transverse weightings both at the level of matrix elements and for the TMDs allows for matching TMDs with matrix elements, by looking at the behavior under time reversal symmetry and the rank of those objects. 

As can be seen from Eq.~\ref{e:GluonCorr}, TMDs with no prefactors of $p_{\scriptscriptstyle T}$ survive direct integration over $p_{\scriptscriptstyle T}$. For TMDs with a prefactor of $p_{\scriptscriptstyle T}^{\alpha}$ one has to use a single transverse weighting, for TMDs with a prefactor of $p_{\scriptscriptstyle T}^\alpha p_{\scriptscriptstyle T}^\beta$ one has to use a double transverse weighting to extract it, etc. The number of $p_{\scriptscriptstyle T}$'s needed for the extraction is called the rank of the TMD and is equal to the number of partonic operators in the matrix elements described before. By using the rank and the behavior under time reversal symmetry one can identify which matrix elements and TMDs correspond to each other. E.g. the correlator $\widetilde\Gamma_{\partial}^{\alpha}(x)$ corresponds to $g_{1T}^g$, since they are both rank 1 and T-even.

\section{Defining gluon TMDs}
The procedure of taking transverse moments can be generalized up to all orders, by including operators up to rank 3, resulting in the expression
\begin{eqnarray}
\Gamma^{[U]}(x,p_{\scriptscriptstyle T}) &\ =\ &
\Gamma(x,p_{\scriptscriptstyle T}^2) 
+ \frac{p_{{\scriptscriptstyle T} i}}{M}\,\widetilde\Gamma_\partial^{i}(x,p_{\scriptscriptstyle T}^2)
+ \frac{p_{{\scriptscriptstyle T} ij}}{M^2}\,\widetilde\Gamma_{\partial\partial}^{ij}(x,p_{\scriptscriptstyle T}^2)
+ \frac{p_{{\scriptscriptstyle T} ijk}}{M^3}\,\widetilde\Gamma_{\partial\partial\partial}^{\,ijk}(x,p_{\scriptscriptstyle T}^2) 
+\ldots 
\nonumber \\ &\quad +&
\sum_c C_{G,c}^{[U]}\left\lgroup\frac{p_{{\scriptscriptstyle T} i}}{M}\,\Gamma_{G,c}^{i}(x,p_{\scriptscriptstyle T}^2)
+ \frac{p_{{\scriptscriptstyle T} ij}}{M^2}\,\widetilde\Gamma_{\{\partial G\},c}^{\,ij}(x,p_{\scriptscriptstyle T}^2)
+ \frac{p_{{\scriptscriptstyle T} ijk}}{M^3}\,\widetilde\Gamma_{\{\partial\partial G\},c}^{\,ijk}(x,p_{\scriptscriptstyle T}^2) + \ldots\right\rgroup
\nonumber \\ &\quad +&
\sum_c C_{GG,c}^{[U]}\left\lgroup\frac{p_{{\scriptscriptstyle T} ij}}{M^2}\,\Gamma_{GG,c}^{ij}(x,p_{\scriptscriptstyle T}^2)
+ \frac{p_{{\scriptscriptstyle T} ijk}}{M^3}\,\widetilde\Gamma_{\{\partial GG\},c}^{\,ijk}(x,p_{\scriptscriptstyle T}^2)+ \ldots\right\rgroup
\nonumber \\ &\quad +&
\sum_c C_{GGG,c}^{[U]}\left\lgroup\frac{p_{{\scriptscriptstyle T} ijk}}{M^3}\,\Gamma_{GGG,c}^{ijk}(x,p_{\scriptscriptstyle T}^2)
+ \ldots \right\rgroup + \ldots \, ,
\label{e:TMDstructure}
\end{eqnarray}
where $p_{{\scriptscriptstyle T} ij}$ and $p_{{\scriptscriptstyle T} ijk}$ are symmetric and traceless tensors and the index $c$ labels the allowed color structures.
Using the behavior under time reversal symmetry and the rank of the operators, we can invert Eq.~\ref{e:TMDstructure} to find which matrix element corresponds to which TMD. Going one step further, one could write down expressions for the TMDs in terms of all the allowed universal functions,
\begin{eqnarray}
f_{1T}^{\perp g[U]}(x,p_{\scriptscriptstyle T}^2)&=&\sum_{c=1}^2 C_{G,c}^{[U]}\,f_{1T}^{\perp g(Ac)}(x,p_{\scriptscriptstyle T}^2), \\
h_{1T}^{g[U]}(x,p_{\scriptscriptstyle T}^2)&=&\sum_{c=1}^2 C_{G,c}^{[U]}\,h_{1T}^{g(Ac)}(x,p_{\scriptscriptstyle T}^2), \\
h_{1L}^{\perp g[U]}(x,p_{\scriptscriptstyle T}^2)&=&\sum_{c=1}^2 C_{G,c}^{[U]}\,h_{1L}^{\perp g(Ac)}(x,p_{\scriptscriptstyle T}^2), \\
h_1^{\perp g[U]}(x,p_{\scriptscriptstyle T}^2)&=&h_1^{\perp g (A)}(x,p_{\scriptscriptstyle T}^2)+\sum_{c=1}^{4}C_{GG,c}^{[U]}\,h_1^{\perp g (Bc)}(x,p_{\scriptscriptstyle T}^2), \\
h_{1T}^{\perp g[U]}(x,p_{\scriptscriptstyle T}^2)&=&\sum_{c=1}^2 C_{G,c}^{[U]}\,h_{1T}^{\perp g(Ac)}(x,p_{\scriptscriptstyle T}^2)+\sum_{c=1}^{7}C_{GGG,c}^{[U]}\,h_{1T}^{\perp g(Bc)}(x,p_{\scriptscriptstyle T}^2).
\end{eqnarray}

\section{Conclusions}
Our main result is that a potentially infinite number of TMDs can be described by a finite number of universal functions. This is of profound importance, since it restores predictability. By writing the (process dependent) TMDs as a sum over a finite number of universal TMDs, one can use experimental results from a few processes to describe the TMDs in any other process. Another important result is that also T-even TMDs can be process dependent, as is the case for $h_1^{\perp g}$. This was shown before for the Pretzelocity (quark) TMD. A more extended account of these results can be found in Ref.~\cite{Buffing:2013kca}.

\acknowledgments
This research is part of the research program of the ``Stichting voor Fundamenteel Onderzoek der Materie (FOM)'', which is financially supported by the ``Nederlandse Organisatie voor Wetenschappelijk Onderzoek (NWO)''. Also support of the FP7 EU-programme HadronPhysics3 (contract no 283286) and QWORK (contract 320389) is acknowledged.
AM thanks the Alexander von Humboldt Fellowship for Experienced Researchers, Germany, for support.


\end{document}